\input{aipcheck}

\documentclass[
    ,final            
  ]
  {aipproc}

\layoutstyle{6x9}

\begin{document}

\title{Muon Lifetime and Muon Capture}

\author{Bernhard Lauss\newline 
{\it\small on behalf of the MuCAP\cite{MuCAP} 
and MuLAN\cite{MuLAN} Collaborations}}
{address={University of California at Berkeley,
Physics Department, 366 LeConte Hall, 
and Lawrence Berkeley National Laboratory,
Berkeley, CA, 94720, USA}
}

\begin{abstract}
We present an introduction to the
MuLAN and MuCAP experiments at PSI, which aim at
high precision determinations
of two fundamental Weak Interactions parameters:
the Fermi constant $G_{F}$ and
the induced pseudoscalar form factor $g_{p}$, respectively.
\end{abstract}

\maketitle

\section{MuLAN - Muon Lifetime}

The Fermi coupling constant $G_{F}$ is one of the fundamental constants
of the Standard Model. $G_{F}$ is obtained from the 
muon lifetime
via a calculation in the Fermi Model, in which weak interactions
are represented by a contact interaction.

\begin{figure}[h]
\includegraphics[height=.1\textheight]{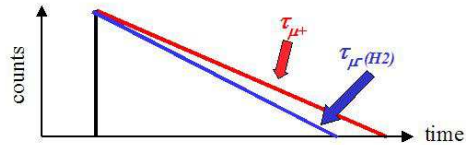}
\caption{The experimental principle of the lifetime method is
to measure the time difference between a muon stopping in 
a target and its decay electron or positron. In the case of $\mu^{+}$
this ideally results in a single exponential 
with $\lambda_{0}=1/ \tau_{\mu^{+}}$ (MuLAN);
for $\mu^{-}$ in hydrogen the rate is increased due 
to the capture process
to be $\lambda=\lambda_{0}+\lambda_{capture}$. The capture rate
consequently follows by comparing the lifetimes of both 
muon charge states in hydrogen (MuCAP).
}
\end{figure}

The goal of the MuLAN ({\bf Mu}on {\bf L}ifetime {\bf An}alysis) 
experiment \cite{www-mulan} is the determination
of the positive muon lifetime, $\tau_{\mu^{+}}$, 
with a precision of 1~ppm; the method is sketched in Fig.1.
In order to achieve this higher 
precision than all combined existing experimental results,
the statistics and systematics
of the measurement have to be dramatically improved.

The ingredients to achieve this challenging goal are:
1) Construction of a multisegment (170 tiles) detector, read out
via fast (500~MHz, 8 bit) waveform digitizers; both are crucial in order to 
separate pile-up events of two simultaneous decay electron hits
in one detector module. The layout of the detector and its elements 
is shown in Fig.2. A depolarizing and dephasing target material -
e.g. sulfur - will be used in a 70~Gauss magnetic field to control
the muon spin rotation from residual muon polarization. 

2) Construction of a new kicked muon beam line in the high
intensity $\pi E3$ muon channel at the Paul Scherrer Institute (PSI). 
This will allow us to collect $10^{12}$ events in a few weeks.
Muons will be stored in the target in a $\sim 5 \mu s$ period 
followed by a 22$\mu s$ ($10 \times \tau_{\mu}$) detection period with
the electrostatic kicker off. 

In a recent run we successfully
installed and tested the kicker in the $\pi E3$ beam line,
which demonstrated the feasibility of the measurements timing.
A offline run with the detector is under way to 
test the full setup with LEDs simulating hits. Fall'03 will see
a first beam test.

\begin{figure}
\includegraphics[height=.3\textheight]{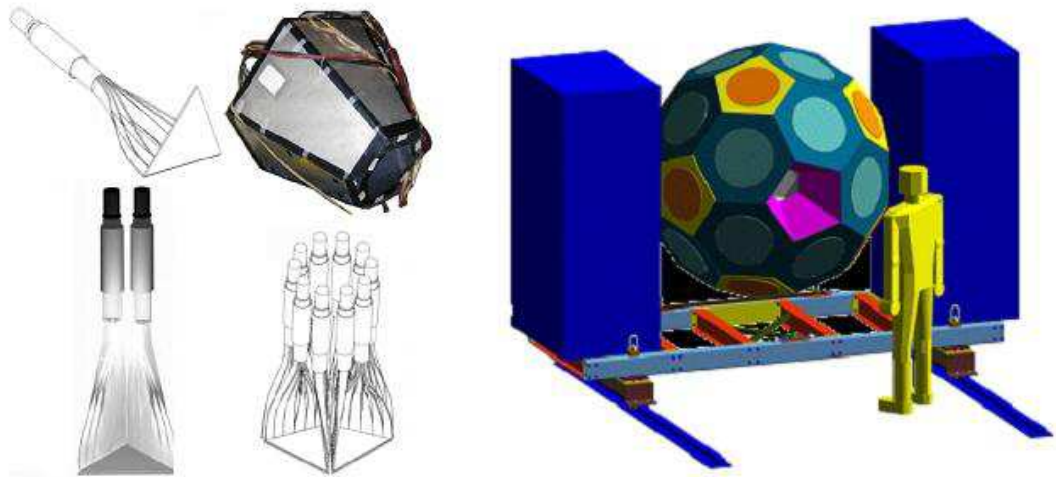}
\caption{Individual scintillation counters, 
double layered scintillator tiles;
their arrangement to a detector module;
schematic view of the full {\bf MuLAN}
soccer-ball detector and electronic racks.
}
\end{figure}

\section{MuCAP - Muon Capture}

\begin{figure}[h]
\includegraphics[height=.28\textheight]{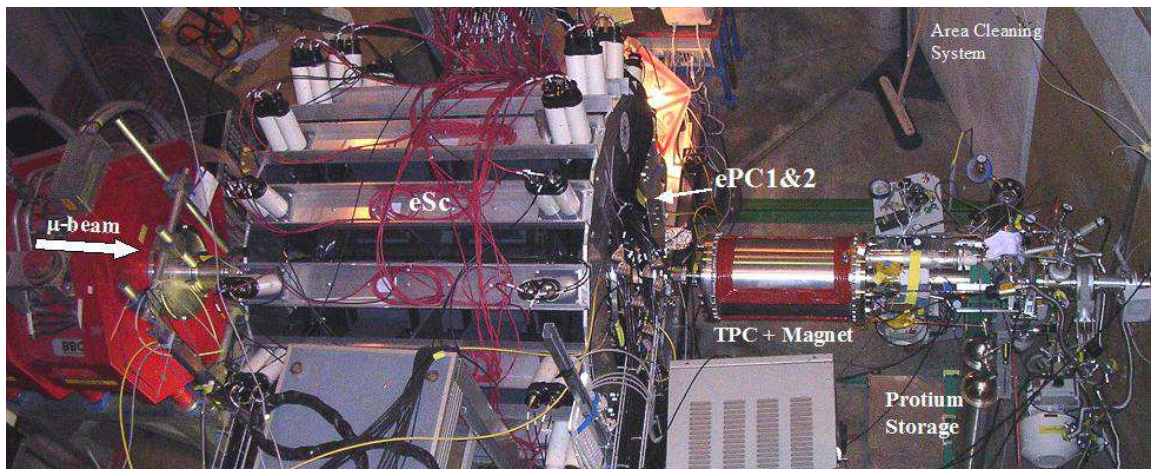}
\caption{View of the 
{\bf MuCAP} experiment located in PSI's
$\mu E4$ area. From left to right: 
the final beamline quadrupoles;
the cylindrical scintillator hodoscope (eSc) with photomultiplier
tubes;
the TPC with the surrounding magnet
rolled back from its center position inside the hodoscope;
the hydrogen purification and filling apparatus.
}
\end{figure}

The goal of MuCAP\cite{www-mulan} is
the high precision measurement of the singlet 
{\bf Mu}on {\bf Cap}ture rate on the proton in low-density, 
ultra-clean $H_{2}$ gas.
The muon capture rate on the proton $\lambda_{cap}$ can be directly
related to the induced pseudoscalar form factor $g_{p}$
of the nucleus,
which was calculated very 
accurately with recent heavy baryon
chiral perturbation theory approaches. 
Existing experimental data are outdated, lack 
accuracy, and show a unresolved discrepancy between
results from ordinary muon capture 
and radiative muon capture.
MuCAP determines the capture rate via the lifetime method
as sketched in Fig.1. Goal of MuCAP is a 1$\%$ precision on
$\lambda_{cap}$, which in turn yields a 7$\%$ error on $g_{p}$.
This is very challenging, due to the large difference in involved rates,
($\lambda_{cap} \sim 700 s^{-1}$,
$\lambda_{0}$=$455000 s^{-1}$,
$\lambda_{transfer~to~Z>1} \sim 10^{9} s^{-1}$
),
and due to the complex chemistry of negative 
muons in hydrogen.

\begin{figure}[h]
\includegraphics[height=.47\textheight]{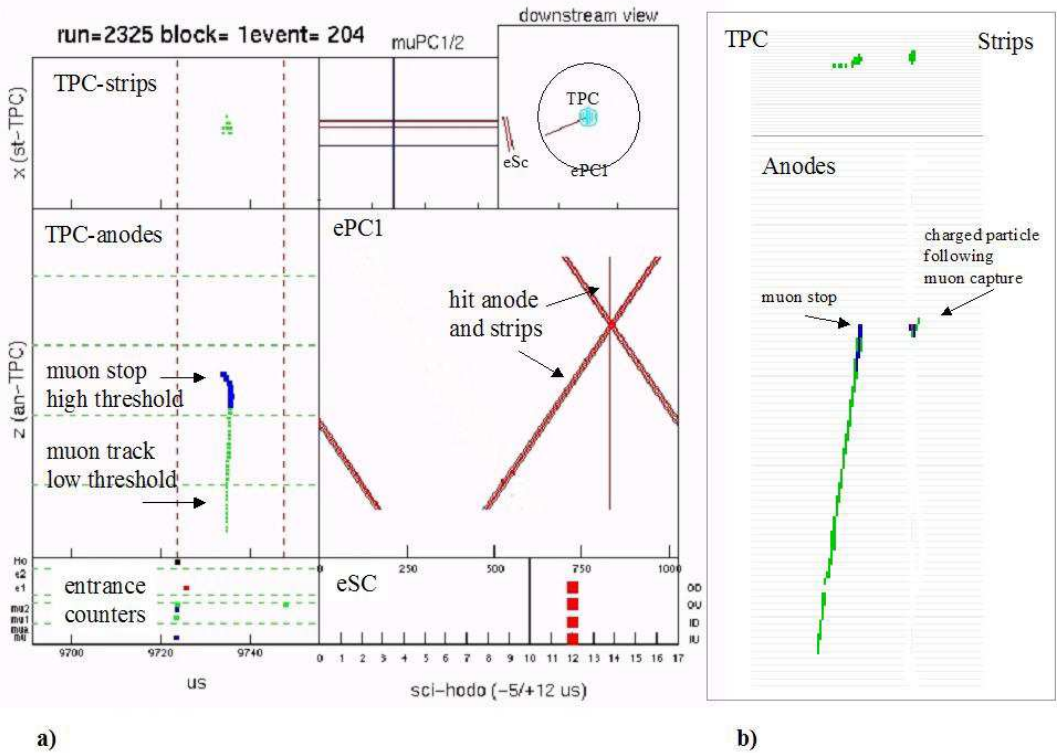}
\caption{a) Typical MuCAP event: Hits in the entrance scintillator
(mu) and wire chambers (muPC1/2) are followed by a muon stop 
in the TPC. The stopping muon triggers a higher 
threshold as it deposits more energy at the end of the Bragg range.
The dashed lines demonstrate the allowed range of 24 $\mu s$ drift time in
the TPC after the initial entrance scintillator hit.
The muon decay electron is observed in one wire 
chamber (ePC1) and in the four-fold hodoscope coincidence (eSC).
b) Detected impurity capture event in the TPC, 
most likely on a nitrogen nucleus.
The muon stops and after a short time a very large signal from a 
charged particle occurs. The time difference is defined by the
muon transfer rate times the impurity concentration.
}
\end{figure}

As part of the effort to control the molecular processes, 
ultra-clean target conditions are selected which enhance
only muonic atomic singlet states and
suppress muonic molecular formation.
A unique high pressure (10 bar) pure hydrogen time projection
chamber (TPC) serves as an active target 
detector. To meet stringent
purity conditions it is made out of quartz-glass and 
bakeable up to 130 degree C.
It is surrounded by a 
$\mu SR$-controlling saddle-coil magnet which 
provides a 70~Gauss field (relevant only for the 
$\mu^{+}$ measurement, as negative muons are effectively
depolarized in the atomic cascade);
two large cylindrical wire chambers (ePC1/2);
and a 16-tile, two-layer scintillator hodoscope (eSc), 
which sees approximately 2/3 of all decay electrons 
in coincidence. A view of the 
setup is presented in Fig.3. 
The timing start with a hit in the entrance scintillator
(mu) and stops with a hit in the scintillator hodoscope
counters,
both with excellent timing resolution. The eSc is
read out with fast waveform digitizers.

The TPC is essential for control of the systematics, 
for several reasons: 
1) It allows the unambiguous identification in 3D of 
the muon stopping positions in hydrogen (and 
consequently excludes wall stops).
2) It can detect impurity captures
(our high Z contamination in the hydrogen after passing the 
palladium filter is smaller than 0.1~ppm and can actively be
monitored via muon capture events on the 
contaminant nuclei - Fig.4b).
3) Muon transfer to deuterium can be observed:
A mismatch between muon stopping position
and back-tracked decay electron
in the two surrounding wire chambers
indicate $\mu d$ diffusion events. 

Fig.4a shows a typical event where a muon, after being seen
in all entrance counters, stops in the TPC.
Most of the time muons only trigger 
a low threshold, but in the final part of the track, where 
there is a large energy deposition, the high threshold
is also fired.
The decay electron is observed in surrounding counters.
Fig.4b shows a detected impurity event: A stopping  
muon followed shortly after by a very high
threshold trigger.

As of summer'03 a MuCAP commissioning run is in progress
and first physics data are expected soon.

\bibliographystyle{aipprocl}

\end{document}